\documentclass[prc,aps,twocolumn,nofootinbib,superscriptaddress,showpacs]{revtex4-1}
\usepackage{graphicx}
\usepackage{amssymb}
\usepackage{amsmath}
\usepackage{dcolumn}
\usepackage{bm}
\begin{document}

\title{Neutrino scattering in supernovae and the universal spin correlations of a unitary gas}

\author{Zidu Lin}
\affiliation{Center for Exploration of Energy and Matter and Department of Physics, Indiana University, Bloomington, IN 47408, USA}

\author{C.\ J.\ Horowitz}
\email[E-mail:~]{horowit@indiana.edu}
\affiliation{Center for Exploration of Energy and Matter and Department of Physics, Indiana University, Bloomington, IN 47408, USA}
%\author{O.\ L.\ Caballero}
%\affiliation{Department of Physics, University of Guelph, Guelph, Ontario N1G 2W1, Canada}
%\author{Evan O'Connor}
%\affiliation{Department of Physics, North Carolina State University, Raleigh, NC 27695, USA; Hubble  Fellow}
%\author{A. Schwenk}
%\affiliation{Institut f\"ur Kernphysik, Technische Universit\"at Darmstadt, 64289 Darmstadt, Germany}
%\affiliation{ExtreMe Matter Institute EMMI, GSI Helmholtzzentrum f\"ur Schwerionenforschung GmbH, 64291 Darmstadt, Germany}
%\affiliation{Max-Planck-Institut f\"ur Kernphysik, Saupfercheckweg 1, 69117 Heidelberg, Germany}
\date{\today}
\begin{abstract}
Core collapse supernova simulations can be sensitive to neutrino interactions near the neutrinosphere.  This is the surface of last scattering.  We model the neutrinosphere region as a warm unitary gas of neutrons.  A unitary gas is a low density system of particles with large scattering lengths.  We calculate modifications to neutrino scattering cross sections because of the universal spin and density correlations of a unitary gas.  These correlations can be studied in laboratory cold atom experiments.  We find significant reductions in cross sections, compared to free space interactions, even at relatively low densities.  These reductions could reduce the delay time from core bounce to successful explosion in multidimensional supernova simulations. 
\end{abstract}

%\pacs{21.65.+f, 26.50.+x, 25.30.Pt, 97.60.Bw}

\maketitle 

%\section{Introduction}

Neutrinos radiate $99 \%$ of the energy and play a crucial role in
core-collapse supernovae~\cite{SNJanka,SNBurrows,SNMezza}. The
scattering of neutrinos and their transport of energy to the shock
region are sensitive to the physics of low-density nucleonic matter,
which is a complex problem due to the strong
correlations induced by nuclear forces. A recent three-dimensional
supernova simulation was sensitive to modest changes in
neutral-current neutrino-nucleon interactions and exploded when
strange-quark contributions were included~\cite{Garching3D}. However,
these strange-quark contributions were probably taken to be
unrealistically large~\cite{strange}. In a recent paper \cite{CJH2017}, we found that similar reductions in neutral-current interactions can arise, not from strange-quark contributions but, from correlations in low-density
nucleonic matter.   Recent two-dimensional supernova simulations find that these reductions of neutrino interactions, from correlations, can impact supernova dynamics and may reduce the delay time from core bounce to successful explosion \cite{Burrows2016, Oconnor2017}, see also \cite{Muons}.   
Note that the physics of neutrino-matter interactions is a
broad and active field, where many interesting studies of
neutrino-matter interactions have been performed over the years, see for example~\cite{Horowitz1991,BS,RPLP,Horowitz2000,Horowitz2004,Burrows2006,Horowitz2012,Bacca2012,Fischer2013,Bartl2014,Rrapaj2015,Balasia2015,Sharma2015,Fischer2016,Bartl2016, Reddy2017}.  Furthermore, we have modeled both neutron and nuclear matter in a virial approximation \cite{vEOSnuc,vEOSneut} and used this to calculate neutrino interactions \cite{CJH2017,response,mass3,light}.

Neutrinos decouple from matter near the neutrinosphere.  Here the details of neutrino interactions can be particularly important for supernova simulations.  The neutrinosphere region is typically a warm, {\it low density} gas of neutron rich matter at densities near $10^{12}$ g/cm$^3$.  At these low densities, around 1/100 of nuclear density, a typical distances between neutrons is of order 8 fm.  This distance is both smaller than the very large neutron-neutron scattering length $a_{nn}\approx -19$ fm and larger than the neutron-neutron effective range $r_0=2.8$ fm \cite{nnscatteringlength}.   Note that the S-wave phase shift $\delta$ at low energies, or wave number $k$, is expanded
%\begin{equation}
$k\cot\delta = -1/a_{nn}+\frac{1}{2}r_0k^2+O(k^4).
$
%\end{equation} 

A unitary gas is a system where the scattering length is infinite $|a_{nn}|\rightarrow \infty$ and the effective range is near zero $r_0\rightarrow 0$.   Because of the large nn scattring length, and the low density, matter near the neutrinosphere should approximate well a warm unitary gas.   {\it This is important because unitary gasses are universal}.  Any system with large scattering length and short effective range should behave in the same way.  
Several unitary gasses of cold atoms have been studied in the laboratory.  
%If one considers a nuclear gas as a unitary gas, in particular, any nucleon-nucleon potential with similar $a_{nn}$ and $r_0$ should give similar results for neutrino scattering.

%For low densities and high temperatures, the virial expansion provides
%a model-independent approach. In previous works, we have presented the
%virial equation of state of low-density nuclear matter~\cite{vEOSnuc}
%and of pure neutron matter~\cite{vEOSneut}. In particular, the virial
%expansion can be used to describe matter in thermal equilibrium around
%the neutrinosphere in supernovae. The temperature of the
%neutrinosphere is roughly $T \sim 4$~MeV from about $20$ neutrinos
%detected in SN1987a~\cite{sn1987a1,sn1987a2} and the mass density is $\rho
%\sim 10^{11}-10^{12} \, \text{g}/\text{cm}^{3}$. For pure neutron
%matter, the virial expansion in terms of the fugacity $z = e^{\mu/T}$
%is valid for
%\begin{equation}
%\rho = \frac{2m}{\lambda^3} \, z + {\mathcal O}(z^2) \, \lesssim \,
%4 \times 10^{11} \, (T/\text{MeV})^{3/2} \, \text{g}/\text{cm}^{3} \,,
%\label{zneutmatt}
%\end{equation}
%where $m$ is the nucleon mass and $\lambda = (2\pi/mT)^{1/2}$ denotes the thermal wavelength. A
%conservative validity range of the virial equation of state is given
%by $z < 1/2$, which gives the limiting density in
%Eq.~(\ref{zneutmatt}). Therefore, the virial approach is applicable
%for the conditions of the neutrinosphere. Following our virial
%equation of state, we have generalized the approach to study
%spin-polarized neutron matter and the consistent neutrino response of
%neutron matter at low densities~\cite{response}.

In this paper, {\it we model the neutrinosphere region as a unitary gas}.  We believe this is a better approximation than modeling the neutrinosphere as a free Fermi gas, as is often done in core collapse supernova simulations.   There are many theoretical calculations of properties of a unitary gas.  In particular, we are interested in neutrino interactions with a unitary gas.  Neutrinos have large spin couplings (from the axial current) to nucleons.  Therefore, we are most interested in the spin response of a unitary gas.  This function describes correlations between the spins of particles in the gas and provides the linear response of the system to any weakly interacting probe that couples to spin. 

It is very important that one can study systems of cold atoms, with large scattering lengths, in the laboratory.  This allows one to experimentally verify properties of unitary gases.  In contrast, it can be difficult to directly study a warm neutron gas.  We will discuss some present cold atom experiments and suggest future cold atom experiments that could measure properties directly relevant for the supernova neutrinosphere.

%use the virial expansion to describe how neutrinos
%interact with low-density nuclear matter composed of protons and
%neutrons. We neglect alpha particles and other light
%nuclei~\cite{mass3,light}. These will be included in later work. 
%In Sec.~\ref{response}, we present our formalism. Our results for the
%axial response and preliminary one-dimensional supernova simulations
%are presented in Sec.~\ref{results}. Finally, we conclude in
%Sec.~\ref{conclusions}.

%\section{Neutrino response}
%\label{response}

First we describe how neutrinos interact with a warm unitary gas.  We focus on neutrino neutral-current
interactions.  These are an important opacity source for mu and tau neutrinos in a supernova.  We expect similar results for charged-current reactions, however we leave these to later work.  Next we will use 
a virial expansion to describe properties of a warm unitary gas and how these modify neutrino interactions in the medium. The virial expansion provides model-independent results for neutrino interactions in the limit of low momentum transfer $q\rightarrow 0$.

The free cross section for neutrino-neutron neutral-current scattering is
\begin{equation}
\frac{d\sigma_0}{d\Omega}=\frac{G_F^2 E_\nu^2}{16\pi^2} \Bigl(g_a^2
(3-\cos\theta)+ 1+\cos\theta \Bigr) \,,
\label{sigmanuN}
\end{equation}
where $G_F$ is the Fermi constant, $E_\nu$ the neutrino energy, and
$\theta$ the scattering angle. The axial coupling constant is $|g_a|= 1.26$.  The cross section in Eq.~(\ref{sigmanuN}) neglects corrections of order $E_\nu/m$, with $m$ the nucleon mass.  These corrections arise from weak magnetism and other effects, for details see~\cite{weakmag}.

%The free cross section per unit volume for scattering from a mixture of
%neutrons and protons is then given by
%\begin{align}
%\frac{1}{V} \frac{d\sigma_0}{d\Omega} &= n_n \, \frac{d\sigma_0}{d\Omega}_{\nu n}+ n_p \, \frac{d\sigma_0}{d\Omega}_{\nu p} \,, \\[2mm]
%&= \frac{G_F^2 E_\nu^2}{16\pi^2} \Bigl( g_a^2 (3-\cos\theta)(n_n+n_p) \nonumber \\
%& \quad\quad\quad\quad + (1+\cos\theta)n_n \Bigr) \,. \label{sigma0}
%\end{align}

In the medium this cross section is modified by the density (vector)
$S_V$ and the spin (axial) $S_A$ response functions.  The response of the
system to density fluctuations is described by $S_V$, while $S_A$
describes the response of the system to spin fluctuations.  
%The response functions are normalized to unity in the low-density limit $S_V, S_A \rightarrow 1$ as $n \rightarrow 0$.  
The cross section per nucleon, in the medium, is then given by
\begin{equation}
\frac{d\sigma}{d\Omega} = \frac{G_F^2 E_\nu^2}{16\pi^2} 
\Bigl(g_a^2(3-\cos\theta)S_A +(1+\cos\theta)S_V \Bigr) \,.
\label{sigma}
\end{equation}
%Note that \$d^2\sigma/d\Omega d\omega$ reduces to the free cross section $d\sigma_0/d\Omega$ as $S_A, S_V \rightarrow 1$. 

Neutrinos interact very weakly with matter.  Therefore, the cross section for neutrino scattering follows from linear response theory involving $S_A(q,\omega)$ and $S_V(q,\omega)$.  In general, these dynamical response functions depend on the momentum transferred from the neutrino to the nucleons $q$ and on the energy transferred $\omega$. Both $S_V(q,\omega)$ and $S_A(q,\omega) $have been measured for a unitary gas of cold $^6Li$ atoms using Bragg spectroscopy \cite{unitaryresponse}.  The spin response $S_A(q,\omega)$ is observed to be reduced compared to that of a free Fermi gas, while $S_V(q,\omega)$ shows an additional peak at lower $\omega$ that corresponds to scattering from a correlated pair of atoms.  

These measurements were done at a relatively low temperature $T\approx 0.1 \epsilon_F$, compared to the Fermi energy $\epsilon_F$.  At this temperature the system is in a superfluid state.  In contrast, supernova matter is often much warmer.  Near the neutrinosphere $T\approx 2-3 \epsilon_F$.  At these temperatures the unitary gas is in a normal state.  It would be very useful to have measurements of $S_A(q,\omega)$ and $S_V(q,\omega)$ as in ref. \cite{unitaryresponse} but for larger temperatures (and ideally for lower momentum transfers $q$, see below).  

Often one does not need the full energy information in $S_A(q,\omega)$, but can instead deal with energy integrated static quantities.  In the rest of the paper we focus on $S_V(q)=\int d\omega S_V(q,\omega)$ and $S_A(q)=\int d\omega S_A(q,\omega)$.  In the limit $q\gg k_F$, there are exact results for $S_V(q)$ and $S_A(q)$ valid for any temperature.  The static structure factor $S_V(q)$ for large $q$ involves the Fourier transform of the radial distribution function at short distances.   The Tan contact $I(T/\epsilon_F)$ describe the probability to find two particles within range of the interactions and determines both the universal radial distribution function at short distances and the high momentum tail of the momentum distribution.  For large momentum transfers, $S_V(q\gg k_F)$ and $S_A(q\gg k_F)$ are \cite{paircorrelations},
\begin{equation}
S_V(q\gg k_F) = 1 + \frac{I(T/\epsilon_F)}{4}\frac{k_F}{q}, 
\label{eq.Svgg}
\end{equation}
\begin{equation}
S_A(q\gg k_F) = 1 -\frac{I(T/\epsilon_F)}{4}\frac{k_F}{q}\, .
\label{eq.Sagg}
\end{equation} 
These equations can be directly used to determine the interaction of high energy neutrinos with supernova matter.   %Given the large $3g_a^2$ coefficient in Eq. \ref{sigma} , the reduction in $S_A$ will lead to a net reduction for the in medium cross section, compared to that in free space.
    
However, most neutrinos in supernovae have relatively low energies $E_\nu\approx 3 T$.  These neutrinos scatter with $q\approx E_\nu \ll k_F$.   Therefore we are most interested, not in the $q\gg k_F$ limit, but in the opposite long wavelength limit $q\rightarrow 0$.   In this limit one can derive model independent results from the virial expansion.

%\subsection{Virial equation of state}

We start by reviewing the virial expansion for a unitary gas of, possibly polarized, spin 1/2 fermions \cite{unitaryvirial}.
%~\cite{vEOSnuc}  
We will use this to calculate
$S_V(q\rightarrow 0)$ and $S_A(q\rightarrow 0)$.  The pressure $P$ is expanded in powers of the
fugacities of spin up particles $z_1=\exp(\mu_1/T)$ with chemical potential $\mu_1$ and spin down particles $z_2=\exp(\mu_2/T)$ with chemical potential $\mu_2$,
\begin{equation}
P = \frac{T}{\lambda^3} \sum_{n_1,n_2}b_{n_1,n_2}z_1^{n_1}z_2^{n_2} \,.
\label{eq.v1}
\end{equation}
Here $b_{n_1,n_2}$ is an $n_1+n_2$ order virial coefficient for a system with $n_1$ spin up and $n_2$ spin down particles.  We will work to fourth order $n_1+n_2\le 4$.
Finally, $T$ is the temperature, and $\lambda=[2\pi/(mT)]^{1/2}$ is the thermal wavelength of particles of mass $m$.  The virial coefficients for a noninteracting spin 1/2 Fermi gas are $b^0_n=b_{n,0}=(-1)^{n+1}/n^{5/2}$, see Table \ref{table1}.  Note that $b_{n_1,n_2}=b_{n_2,n_1}$ and, for a unitary gas, there are no interactions between like spin particles.    

For an unpolarized gas, $z_1=z_2=z$, Eq. \ref{eq.v1}. reduces to,
\begin{equation}
P=\frac{2T}{\lambda^3}\sum_{n=1}^4b_n z^n
\label{eq.v2}
\end{equation}
with $b_1=1$, $b_2=b_2^0+b_{1,1}/2$, $b_3=b_3^0+b_{2,1}$, and $b_4=b_4^0+b_{3,1}+b_{2,2}/2$.  The values of these virial coefficients are collected in Table \ref{table1}.  Our conventions are to include the noninteracting contributions $b_n^0$ in $b_n$ and we note that all of our virial coefficients are for a uniform infinite system rather than a harmonic trap.  The density of the system $n$ is 
\begin{equation}
n=\frac{z}{T}\frac{dP}{dz}=\frac{2}{\lambda^3}\bigl[z+2z^2b_2+3z^3b_3+4z^4b_4\bigr]\, .
\end{equation}
For this density the Fermi momentum is $k_F=[(3\pi^2)n]^{1/3}$ and we define a Fermi energy $\epsilon_F=k_F^2/(2m)$ so that the degree of degeneracy is related to $\epsilon_F/T$,
\begin{equation}
\frac{\epsilon_F}{T}=\bigl(\frac{9\pi}{16}\bigr)^{1/3}\bigl(z+2z^2b_2+3z^3b_3+4z^4b_4\bigr)^{2/3}\,.
\label{eq.eFT}
\end{equation}
For the unitary gas the virial coefficients are independent of temperature so all properties are only functions of $\epsilon_F/T$ instead of depending on $n$ and $T$ separately.  Equation \ref{eq.eFT} can be inverted to find $z$ as a function of $\epsilon_F/T$. 

%\subsection{Vector response}

\begin{figure}[t]
\centering
\includegraphics[width=\columnwidth]{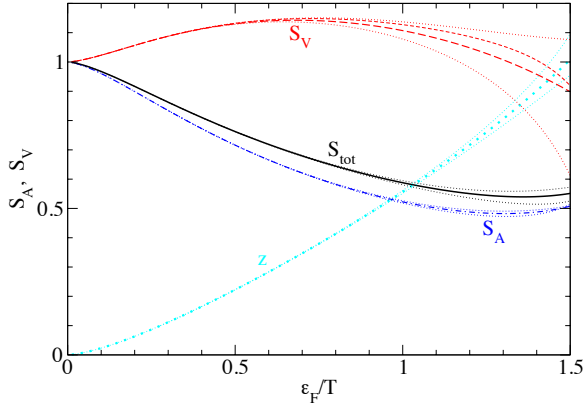}
\caption{(Color online) Fourth order virial calculations of the vector response $S_V$ (red heavy dashed line), axial response $S_A$ (heavy blue dashed-dotted line), and total response $S_{tot}$ (heavy black line) versus Fermi energy over temperature $\epsilon_F/T$.  The fugacity $z=\exp(\mu/T)$ is shown as the heavy Cyan dotted line.  The dotted error bands show the effect of statistical and systematic errors in theoretical calculations of the fourth virial coefficient $b_4$.  Finally the thin red dashed line shows $S_V$ calculated using an approximate virial expansion to tenth order \cite{b10}.  }
\label{fig.1}
\end{figure}

The vector response $S_V$, in the long wavelength limit, can be calculated from the virial equation of state
$S_V(q\rightarrow 0) = T/(\partial P/\partial n)_T = z( \partial n/\partial z)/n$,
\begin{equation}
S_V(q\rightarrow 0) = \frac{1+4zb_2+9z^2b_3+16z^3b_4}{1+2zb_2+3z^2b_3+4z^3b_4} \,.
\label{Svfinal}
\end{equation}
Figure \ref{fig.1} shows $S_V$.  This first increases with $\epsilon_F/T$ because of density fluctuations and then decreases at higher densities because of Pauli blocking.  We emphasize that $S_V$ (and $S_A$) include corrections (contained in $b_n^0$) from the Pauli blocking of the scattered nucleon.

\begin{table}
\begin{tabular}{ccccc}
$n$ & $b_n$ & $b_n^0$ & $b_{3,1}$& Ref.\\
\hline
2 & 0.53033 & -0.17678 & & \\
3 & -0.29095 & 0.06415 & & \cite{b3}\\
4 &  0.047(18) & -0.03125 & 0.170(13) & \cite{unitaryvirial}
\end{tabular} 
\caption{Virial coefficients $b_n$ for a unitary gas, while $b_n^0$ are virial coefficients for a free Fermi gas.  Finally $b_{3,1}$ is the fourth order coefficient for three spin up and one spin down particles (see text).  The numbers in parentheses are the theoretical errors in the fourth order coefficients \cite{unitaryvirial}.} 
\label{table1}
\end{table}

%\subsection{Axial response}

The axial or spin response $S_A$, in the long wavelength limit, can be calculated from the virial equation of state for a spin polarized system, see for example \cite{BS},
\begin{equation}
S_A(q\rightarrow0)=\frac{2z}{n}\frac{\partial}{\partial(z_1-z_2)} (n_1-n_2)\bigr|_{z_1=z_2}\, .
\end{equation}
where $n_i=z_i(dP/dz_i)/T$.  Using Eq. \ref{eq.v1} we get,
\begin{equation}
S_A(q\rightarrow 0)=\frac{1+4zb_2^0+z^2(8b_3^0+b_3)+z^3(16b_4^0+4b_{3,1})}{1+2zb_2+3z^2b_3+4z^3b_4}\,.
\label{eq.Sa}
\end{equation}
Two particles are correlated in the $^1S_0$ state.  This spin zero state reduces the spin response so that $S_A<1$.  This is shown in Fig. \ref{fig.1}.

To summarize, the neutrino cross section in the medium is given by Eq. \ref{sigma} with $S_V$ given by Eq. \ref{Svfinal} and $S_A$ given by Eq. \ref{eq.Sa}.
We define the total response $S_{\rm tot}$ as the ratio of the
in-medium transport cross section to the free one,
\begin{equation}
S_{\rm tot}=\frac{\int d\Omega \, \frac{d\sigma}{d\Omega} (1-\cos\theta)}{\int d\Omega \, \frac{d\sigma_0}{d\Omega} (1-\cos\theta)} =\frac{5g_a^2S_A+S_V}{5g_a^2+1}\,.
\label{eq.Stot}
\end{equation}
Thus $S_{tot}$ is a combination of $S_A$ and $S_V$ and is dominated by $S_A$ because of the large $5g_a^2$ coefficient.

We now discuss the convergence of our virial results and their sensitivity to errors in the virial coefficients.  We use the Path Integral Monte Carlo (PIMC) results for the fourth order coefficients $b_4$ and $b_{3.,1}$ \cite{unitaryvirial}, rather than the somewhat more accurate experimental value for $b_4$ \cite{expb4}, because the PIMC calculations also include a value for $b_{3,1}$ that we need to calculate $S_A$. Figure \ref{fig.1} includes dotted error bands for $S_V$, $S_A$, and $S_{tot}$ obtained by changing $b_4$ and $b_{3,1}$ by their theoretical errors.  We see that $S_A$ and $S_{tot}$ are relatively insensitive.  In contrast, $S_V$ does depend sensitively on $b_4$ for $\epsilon_F/T>1$.\footnote{Note the lower error band for $S_V$ in Fig. \ref{fig.1} may be unrealistic because the experimental value of $b_4$ is close to the value for the upper error band \cite{expb4}.}  Therefore, the convergence of $S_V$, as a function of $z$, may be poorer than the convergence of $S_A$.  This arrises because $S_V$ involves two derivatives of the pressure with respect to $z$.  To test the convergence of the virial expansion for $S_V$ we evaluate it to tenth order using the approximate higher order virial coefficients from ref. \cite{b10}.  This is shown in Fig. \ref{fig.1} and agrees within errors with our calculation.  Note that the convergence of the virial expansion for the pressure is known to be very good \cite{b10}.    We conclude that the results in Fig. \ref{fig.1} should be reliable.

\begin{figure}[t]
\centering
\includegraphics[width=\columnwidth]{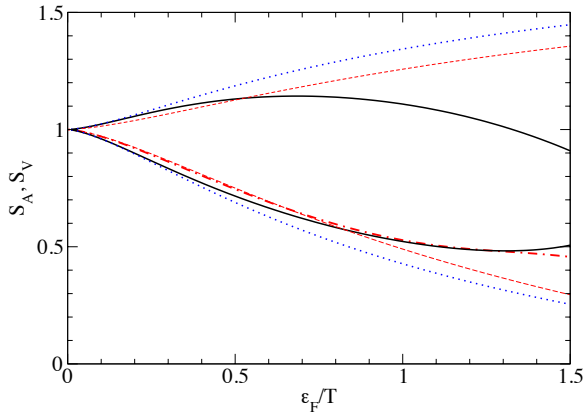}
\caption{(Color online) Vector response $S_V$ (upper three curves) and axial response $S_A$ (lower four curves) versus Fermi energy over temperature $\epsilon_F/T$.  Unitary gas fourth order virial results from Fig. \ref{fig.1} are solid while second order virial results are dotted.  Pure neutron gas results at a temperature of 10 MeV, to second order in the virial expansion, are dashed.  Finally the dot-dashed curve shows the fit from ref. \cite{CJH2017} for pure neutron matter.  This reproduces virial results  at low density and RPA calculations \cite{BS} at high densities.}
\label{fig.2}
\end{figure}

We compare our unitary gas results to earlier virial calculations for pure neutron matter.  We start with the vector response.  Figure \ref{fig.2} shows that neutron matter virial calculations from ref. \cite{CJH2017} are consistent with our unitary gas results for $S_V$ only at low densities.   This is because the neutron matter virial calculations are only to second order.  Indeed second order virial results for the unitary gas are similar to the neutron matter results, but somewhat larger, see Fig. \ref{fig.2}.  This is because the second virial coefficient for a unitary gas $b_2\approx 0.53$ is somewhat larger than for a neutron gas $b_n\approx 0.3$.  Note, $b_2$ for a unitary gas is independent of temperature while $b_n$ for a neutron gas depends very weakly on temperature and we evaluate it at $T=10$ MeV.  We conclude from Fig. \ref{fig.2} that a second order virial calculation may overestimate $S_V$ except at low densities.

We now discuss the axial response.  Figure \ref{fig.2} shows that second order virial calculations do somewhat better jobs of reproducing $S_A$ (than they do for $S_V$).  Finally, ref. \cite{CJH2017} provided a simple fit to $S_A$ that reproduces neutron matter virial results up to a fugacity $\approx 0.5$ and then fits the model dependent RPA calculations of Burrows and Sawyer \cite{BS} at higher densities.  This fit agrees remarkably well with our unitary gas results.  However the present unitary gas calculations are simpler, cleaner, and less model dependent.  Furthermore they can be experimentally verified with laboratory cold atom experiments.

 \begin{table}
\begin{tabular}{ccccc}
\\
$i$ & $z$ & $S_V$ & $S_A$& $S_{tot}$\\
\hline
1.5 & 0.97362 & 1.1153 & -1.7008 & -1.3858 \\
2   & -0.55516 & -1.1148 & +1.3336 & 1.0597\\
3 & 0.13744 & 0.10751 & -0.11221 & -0.08763
\end{tabular} 
\caption{Expansion coefficients $a_i^x$  in fits of $z$, $S_V$, $S_A$, and $S_{tot}$ as a function of $\epsilon_F/T$, see Eq. \ref{eq.fit}.  This fit is valid for $0<\epsilon_F/T<1.5$. }  
\label{table2}
\end{table}

We now consider applying our unitary gas results to astrophysical simulations.  First, we fit the results in Fig. \ref{fig.1} for $S_V$, $S_A$, $S_{tot}$ with the simple functional form
\begin{equation}
S_x\approx 1+a_{1.5}^x (\epsilon_F/T)^{3/2}+a_2^x (\epsilon_F/T)^2 + a_3^x (\epsilon_F/T)^3\, ,
\label{eq.fit}
\end{equation}
for $x=V$, $A$, and $tot$.  This fit is valid for $0<(\epsilon_F/T)<1.5$ and the coefficients $a_i^x$ are given in Table \ref{table2}.  We also fit the fugacity $z\approx a_{1.5}^z(\epsilon_F/T)^{3/2}+a_2^z(\epsilon_F/T)^2+a_3^z(\epsilon_F/T)^3$ in Table \ref{table2}.  

We recommend applying our results to supernova or other astrophysical simulations as follows.  Neutrino-neutron neutral current cross sections are given by Eq. \ref{sigma} with $S_V$ and $S_A$ given by Eq. \ref{eq.fit} and Table \ref{table2}.  Alternatively, one could simply multiply the free-space neutrino-neutron interaction by $S_{tot}$ from Eq. \ref{eq.Stot} and Table \ref{table2}.  We have not explicitly considered small admixtures of protons.  A minimal assumption would be to describe neutrino-proton neutral current scattering by Eq. \ref{sigma} with $S_V=0$, because the weak charge of a proton is small, and $S_A=1$.  We choose $S_A=1$, rather than the reduced unitary gas value, because ref. \cite{CJH2017} finds the reduction in $S_A$ to be somewhat smaller as $Y_e$ increases.   In Eq. \ref{eq.fit} a minimal assumption for $\epsilon_F/T$ is
\begin{equation}
\frac{\epsilon_F}{T}={\rm Min}\bigl\{\frac{(3\pi^2n_n)^{2/3}}{2mT},1.5\bigr\}\, ,
\end{equation}
where $n_n$ is the neutron density.  Our virial results should be valid for $0<\epsilon_F/T<1.5$.  For larger values of $\epsilon_F/T$ we suggest simply using our results evaluated at $\epsilon_F/T=1.5$ as a minimal assumption.  For example, at a temperature of 15 MeV our results are good up to a density of $7\times 10^{13}$ g/cm$^3$.  Neutrino interactions at higher densities may not be very important for supernova dynamics except at later times.   However, we will explore the unitary gas response at higher densities in later work.   Our procedure is based on only the unitary gas response.  However, it should give results similar to the hybrid approach of ref. \cite{CJH2017} that matched on to model dependent RPA results at high densities.  

Future work would be very useful in three areas.  First, calculations of third (or fourth) order virial coefficients for neutron and nuclear matter would be very helpful.  Perhaps this could be done by calculating the energies of three (or four) nucleons in a harmonic trap and taking the limit as the trap frequency goes to zero.  Second, microscopic calculations of the vector and axial responses should be done for both a unitary gas and for neutron and nuclear matter.  These should reproduce our virial results at low densities and be directly applicable at higher densities.  One approach would be to use quasipotentials, that reproduce NN scattering, in a random phase approximation or in many-body perturbation theory.  Finally, more experimental measurements of the dynamical spin response of a unitary gas of cold atoms would be very useful.  These should be done at higher temperatures than pervious measurements \cite{unitaryresponse} and ideally at lower momentum transfers. 

In conclusions, core collapse supernova simulations can be sensitive to neutrino interactions near the neutrinosphere.  In this paper we model the neutrinosphere region as a warm unitary gas of neutrons.   Using the virial expansion to fourth order we calculate modifications to neutrino scattering cross sections because of spin correlations in the unitary gas.  These spin correlations are universal for any unitary gas and can be studied in the laboratory with cold atom experiments.  We find significant reductions in cross sections, even at relatively low densities.  These reductions could reduce the delay time from core bounce to successful explosion in multidimensional supernova simulations.

%\begin{acknowledgments}

We thank Adam Burrows, Jason Ho,  Sanjay Reddy, and Qi Zhou for helpful discussions. This
work was supported in part by DOE Grants DE-FG02-87ER40365 (Indiana
University) and DE-SC0008808 (NUCLEI SciDAC Collaboration).  This work was started at the Aspen Center for Physics, which is supported by National Science Foundation grant PHY-1607611.

%\end{acknowledgments}

\end{document}